\documentclass[hyper]{JHEP3}
\usepackage{amsmath,amssymb}
\usepackage{amsfonts}
\usepackage{verbatim}
\usepackage{latexsym}
\usepackage{amsthm}
\usepackage{amsbsy}
\usepackage{multirow}

\def\one{{\hbox{ 1\kern-.8mm l}}}
\newcommand{\Dslash}{\not{\hbox{\kern-4pt $D$}}}
\newcommand{\cL}{\mathcal{L}}

\newcommand{\Comment}[1]{{}}
\newcommand{\beas}{\begin{eqnarray*}}
\newcommand{\eeas}{\end{eqnarray*}}
\newcommand{\be}{\begin{equation}}
\newcommand{\ee}{\end{equation}}

\def\IZ{{\mathbb Z}}

\def\IR{{\mathbb R}}

\newcommand{\bc}{\begin{center}}
\newcommand{\ec}{\end{center}}
\newcommand{\ba}{\begin{array}}
\newcommand{\ea}{\end{array}}
\newcommand{\beq}{\begin{equation}}
\newcommand{\eeq}{\end{equation}}
\newcommand{\bea}{\begin{eqnarray}}
\newcommand{\eea}{\end{eqnarray}}
\newcommand{\bmx}{\begin{pmatrix}}
\newcommand{\emx}{\end{pmatrix}}
\newcommand{\nn}{\nonumber}

\newcommand{\tr}{{\rm tr}}

\newcommand{\eref}[1]{Eq.~(\ref{#1})}

\newcommand{\tOtwo}{\widetilde{O2}}

\def\IB{\relax{\rm I\kern-.18em B}}
\def\IC{{\relax\hbox{\kern.3em{\cmss I}$\kern-.4em{\rm C}$}}}
\def\ID{\relax{\rm I\kern-.18em D}}
\def\IE{\relax{\rm I\kern-.18em E}}
\def\IF{\relax{\rm I\kern-.18em F}}
\def\II{\relax{\rm I\kern-.18em I}}
\def\Id{\relax{1\kern-.32em 1}}
\def\IG{\relax\hbox{$\inbar\kern-.3em{\rm G}$}}
\def\IR{\relax{\rm I\kern-.18em R}}

\normalsize

\catcode`\@=11
\renewcommand\appendix{\par
  \setcounter{section}{0}%
  \setcounter{subsection}{0}%
  \renewcommand\thesection{\@Alph\c@section}%
  \renewcommand\secstyle{\bfseries\noindent Appendix }}
\catcode`\@=12

\title{M2-branes on M-folds}

\author
{Jacques Distler$^1$,
Sunil Mukhi$^2$,
Constantinos Papageorgakis$^2$ and Mark Van Raamsdonk$^3$
\\ ~\\\it
{}$^1$ Department of Physics,\\
~~ University of Texas, Austin, TX 78712-0264, USA\\ ~\\
{}$^2$ Tata Institute of Fundamental Research,\\ ~~
Homi Bhabha Rd, Mumbai 400
  005, India\\ ~\\
{}$^3$ Department of Physics and Astronomy, University of British
Columbia,\\
~~ 6224 Agricultural Road, Vancouver B.C., V6T 1W9, Canada
}

\abstract{We argue that the moduli space for the Bagger-Lambert
$A_4$ theory at level $k$ is $(\mathbb{R}^8 \times
\mathbb{R}^8)/D_{2k}$, where $D_{2k}$ is the dihedral group of order
$4k$. We conjecture that the theory describes two M2-branes on a
$\IZ_{2k}$ ``M-fold'', in which a geometrical action of $\IZ_{2k}$ is
combined with an action on the branes. For $k=1$, this arises as the
strong coupling limit of two D2-branes on an $O2^-$ orientifold, whose
worldvolume theory is the maximally supersymmetric $SO(4)$ gauge
theory. Finally, in an appropriate large-$k$ limit we show that one
recovers compactified M-theory and the M2-branes reduce to D2-branes.}
\preprint{TIFR/TH/08-13\\ UTTG-03-08}

\begin{document}

\section{Introduction}
\label{Introduction and Review}

A new class of conformal invariant, maximally supersymmetric field
theories in $2+1$ dimensions has been found
recently \cite{Bagger:2007jr,Gustavsson:2007vu}. These theories are
based on ``3-algebras'' and include a non-dynamical gauge field with a
Chern-Simons-like interaction. They have several striking properties
including the absence of continuous marginal deformations. The
motivation for studying these theories was to find a Lagrangian
description of the conformally invariant fixed point of maximally
supersymmetric Yang-Mills theories in $2+1$ dimensions, which is
believed to describe the worldvolume dynamics of coincident membranes
in M-theory.

While the 3-algebra theories share many features with the expected
M2-brane theories, they also give rise to some puzzles. One is that
only a single 3-algebra, denoted $A_4$, is presently known, so an
explicit theory exists for at best a small fixed number of
membranes. It was proposed in Ref. \cite{Bagger:2007vi} that this
number is 3, which suggests the surprising possibility that the IR
theory on 2 D2-branes is trivial. Also somewhat puzzling was how
parity could be preserved when the gauge field has Chern-Simons
interactions.

Some of these puzzles have been resolved in recent
days \cite{Mukhi:2008ux,Bandres:2008vf,VanRaamsdonk:2008ft}. For the
$A_4$ 3-algebra, all these papers (as well as
Ref. \cite{Berman:2008be}) found that the theory could be recast as an
$SU(2)\times SU(2)$ gauge theory. In Ref. \cite{Mukhi:2008ux} it was
further shown that giving one of the scalars a vev reduces the
3-algebra action to a strongly coupled supersymmetric $SU(2)$
Yang-Mills action by a novel Higgs mechanism.  This renders one
combination of the two Chern-Simons fields massive and the other one
dynamical in consequence. In
Refs. \cite{Bandres:2008vf,VanRaamsdonk:2008ft} it was shown that the
theory is parity-invariant if parity is taken to exchange the two
$SU(2)$'s.

However, new puzzles emerged. Ref. \cite{VanRaamsdonk:2008ft} studied
the moduli space of the theory and found that it does not appear to
match expectations for either two or three M2-branes. Additionally the
spectrum of chiral primary operators had some ``missing'' operators
that should have been present for a multiple M2-brane interpretation
to be correct. In this work it was also noted that the level $k$ of
the $SU(2)\times SU(2)$ is a free discrete parameter which, at large
values, causes the theory to become weakly coupled -- but a finite set
of M2-branes should not have any weakly coupled limit. Also,
although Ref. \cite{Mukhi:2008ux} found a result suggestive of
compactification, it was not clear why going to the Coulomb branch
should be related to a circle-compactified background.

In the present work we resolve some of the above puzzles. We
conjecture that the Bagger-Lambert $A_4$ theory at level $k=1$
describes the worldvolume dynamics of two M2-branes on the $\IZ_2$
orbifold, defined by the uplift to M-theory of two D2-branes on an
$O2^-$ orientifold. Equivalently, the level-one $A_4$ theory is the
infrared fixed point of the $SO(4)$ maximally supersymmetric
Yang-Mills theory in 2+1 dimensions. With this interpretation, we
argue that the spectrum of chiral operators is as expected. For
general $k$, we argue that the moduli space is $(\mathbb{R}^8 \times
\mathbb{R}^8)/D_{2k}$, where $D_{2k}$ is the dihedral group of order
$4k$.  We conjecture that this corresponds to two M2-branes on a $\IZ_{2k}$
``M-fold'', in which a geometrical action of $\IZ_{2k}$
is combined with an action on the branes. Finally, we show that taking
a large-$k$ limit at a point on moduli space where the branes are
separated from the orbifold point, one recovers the worldvolume theory
of D2-branes, as expected, since the orbifold locally becomes a
cylinder.

We will work with the formulation of the $A_4$ theory in
Ref. \cite{VanRaamsdonk:2008ft}. The fields consist of two $SU(2)$
gauge fields, having Chern-Simons actions with {\it opposite} signs,
and a set of 8 scalar fields $X^I,I=1,2,\cdots,8$ along with 8
fermions. All the matter fields transform as bi-fundamentals of
$SU(2)\times SU(2)$. The action is:
\bea
\label{markform}
{\cal L} &=&  \tr( -(D^\mu X^I)^\dagger D_\mu X^I +
i \bar{\Psi}^\dagger \Gamma^
{\mu} D_\mu \Psi )\nn\\
&& + \tr(-\frac{2}{3} if
\bar{\Psi}^\dagger \Gamma_{IJ} (X^I X^{J \dagger} \Psi
+ X^J \Psi^\dagger X^I + \Psi X^{I \dagger} X^J)
- \frac{8}{3}f^2 X^{[I} X^{J \dagger} X^{K]}X^{K \dagger}
X^J X^{I \dagger}) \nn\\
&& + \frac{1}{2f} \epsilon^{\mu \nu \lambda}
\tr(A_\mu \partial_\nu A_\lambda +
\frac{2}{3}i A_\mu A_\nu A_\lambda)
- \frac{1}{2f} \epsilon^{\mu \nu \lambda} \
tr(\hat{A}_\mu \partial_\nu \hat{A}_\lambda
+ \frac{2}{3}i \hat{A}_\mu \hat{A}_\nu \hat{A}_\lambda) \; .
\eea
Here,
\beq
D_\mu X^I = \partial_\mu X^I + i A_\mu X^I -i X^I \hat{A}_\mu \; ,
\eeq
which is covariant under the action of the gauge transformations
\beq
X^I \to U X^I V^{-1}, \quad A_\mu
\to U A_\mu U^{-1} + i \partial_\mu U U^{-1},
\quad \hat{A}_\mu \to V \hat{A}_\mu V^{-1}
+ i \partial_\mu V V^{-1} \; .
\eeq
In the above, $f=2\pi/k$ where $k$ is the (integer) level of the two
Chern-Simons actions\footnote{The quantization of the Chern-Simons
coefficient for a non-simply-connected gauge group, $G$, is a little
subtle. See Appendix \ref{SpinCS} for details.}. The supersymmetries under which
the above action is invariant can be found in
Ref. \cite{VanRaamsdonk:2008ft}.

Since our proposal involves orbifold 2-planes, let us briefly review
some relevant facts. There are three types of orientifold 2-planes in
type IIA string theory \cite{Sethi:1998zk}, denoted $O2^-$, $\tOtwo^+$
and $O2^+$, that give rise to gauge groups $SO(2N), SO(2N+1),
Sp(N)$ respectively when $N$ D2-branes are brought near them. All
correspond to an inversion of 7 spatial directions transverse to the
orientifold plane, and all can be uplifted to M-theory. The uplifted
orientifold planes are really M-theory {\it orbifolds} rather than
orientifolds, in the sense that they do not reverse the orientation of
membranes or of the 3-form $C_{MNP}$.\footnote{In contrast,
orientifold 4-planes in type IIA lift to orientifold 5-planes in
M-theory \cite{Dasgupta:1995zm,Witten:1995em}.}  This is because in IIA
string theory, the $\IZ_2$ action reverses the $B_{MN}$ field, but
preserves the RR 3-form $C_{MNP}$. This implies an action on the
M-circle as a reflection. After uplifting, the end result is that it
preserves the 3-form of M-theory but reflects eight spatial directions
including the M-circle. Due to their origin as orientifold planes, the
M2-orbifold planes carry an M2-brane charge, which is $-\frac{1}{16}$
for the $O2^-$ case.

We can directly define the $O2^-$ plane in M-theory as the orbifold
$R^8/\IZ_2$, where the action of $\IZ_2$ is diag $(-1,-1,-1,-1)$ on
the four complex coordinates of $R^8$. With this $\IZ_2$ action the
supersymmetry near the plane is half-maximal and has 16 components
just like the supersymmetry on M2 branes. This will turn out to be the
case we understand best. For $k>1$, the $\mathbb{Z}_2$ subgroup of
$D_2$ associated with the inversion of the $\mathbb{R}^8$ is replaced
with $\mathbb{Z}_{2k}$ in the definition of the moduli space. This
suggests that the level $k$ M-fold combines a geometrical action of
$\IZ_{2k}$ with an action on the branes. While we will not be able to
present a precise action of $\IZ_{2k}$ satisfying all the
requirements, we will discuss some possibilities in a subsequent
section. Unlike $k=1$, the general case is not likely to
descend in a simple way to a type IIA orientifold since a $\IZ_{2k}$
action with $k>1$ will presumably mix the M-circle with another
circle.\footnote{By compactifying a direction transverse to $R^8$ one
can relate it to a type IIA orbifold, however in this case it becomes
an orbifold 1-plane and carries the charge of fundamental strings
rather than D2-branes.}

In the rest of this note, we present evidence for our conjecture that
the theory whose Lagrangian is given in \eref{markform} describes two
$M2$-branes at a $\IZ_{2k}$ orbifold (with the action of $\IZ_{2k}$ on
the brane worldvolume fields appropriately defined).

When the original version of this paper was nearly complete, the paper
\cite{Lambert:2008et} appeared, which has substantial overlap with our work. The original versions of our paper and of \cite{Lambert:2008et} differed in a few significant respects, but most of these differences have now been resolved in the revised versions. We thank David Tong and Neil Lambert for correspondence on these issues. Another very recent paper discussing multiple M2-branes is Ref. \cite{Morozov:2008cb}.

\section{Moduli space}

The moduli space for the $A_4$ theory was studied in
Refs. \cite{Bagger:2007vi,VanRaamsdonk:2008ft}. Here we will revisit
this moduli space and argue that the complete moduli space at
level $k$ is actually $(\mathbb{R}^8 \times \mathbb{R}^8)/D_{2k}$
once the gauge fields are taken into account. Here, $D_{2k}$ is the
dihedral group, $\mathbb{Z}_2 \ltimes \mathbb{Z}_{2k}$ where
the product is semidirect.

We begin with the action in \eref{markform}. As noted in
\cite{VanRaamsdonk:2008ft}, generic scalar configurations for which
the potential vanishes correspond (up to gauge transformations) to
diagonal matrices $X^I$, which we will parameterize by
\be
\label{diag}
X^I = \frac{1}{\sqrt{2}} \left( \ba{cc} z^I
& 0 \cr 0 & \bar{z}^I \ea \right) \; .
\ee
Within the space of these diagonal configurations, there is a residual
$O(2)$ gauge symmetry, acting by simultaneous rotations on $z^I$ and
by simultaneous complex conjugation. However, to describe the complete
moduli space it will be important for us to take into account the
gauge fields.

Generically, the diagonal configurations (\ref{diag}) break the gauge
group down to $U(1)$, and the remaining components of the gauge field
become massive by the Higgs mechanism. Also, expanding the potential
about such configurations shows that physical scalar fluctuations
which take us away from a diagonal configuration are all massive.

We now write the classical action describing the dynamics of the light
fields on the moduli space. To do this, it will be convenient to
include both the preserved $U(1)$ gauge field and the gauge field
associated with the $U(1)$ that rotates $z^I$. Together with the
diagonal configuration (\ref{diag}), we take
\be
A_\mu = \left( \ba{cc} a_\mu & 0 \cr 0 & -a_\mu \ea \right),\qquad
\hat{A}_\mu = \left( \ba{cc} \hat{a}_\mu & 0 \cr 0 & -\hat{a}_\mu
\ea \right)
\ee
with the normalization chosen so that $a_\mu$ and $\hat{a}_\mu$ have
gauge transformations
\be
a_\mu \to a_\mu - \partial_\mu \theta,
\qquad \hat{a}_\mu \to \hat{a}_\mu - \partial_\mu \hat{\theta}
\ee
where $\theta$ and $\hat{\theta}$ have period $2 \pi$. This gives an
action
\be
S = \int d^3 x
\left( - \big| \partial_\mu z^I + i (a_\mu-\hat{a}_\mu) z^I\big|^2
+ \frac{k}{2 \pi} \epsilon^{\mu \nu \lambda}
\big(a_\mu \partial_\nu a_\lambda -
\hat{a}_\mu \partial_\nu \hat{a}_\lambda\big) \right) \; .
\ee
We further define
\be
c_\mu = a_\mu + \hat{a}_\mu, \qquad
b_\mu = a_\mu - \hat{a}_\mu
\ee
so that $c_\mu $ is the gauge field associated with the preserved
$U(1)$, and $b_\mu$ is associated with the $U(1)$ that rotates $z^I$.

The resulting action is
\be
\label{action}
S = \int d^3 x \left( - \big| \partial_\mu z^I + i b_\mu z^I\big|^2
+ \frac{k}{4 \pi} \epsilon^{\mu \nu \lambda}
b_\mu f_{\nu \lambda} \right) \; ,
\ee
where
\be
f_{\mu \nu} = \partial_\mu c_\nu - \partial_\nu c_\mu \; .
\ee
The gauge transformations are
\be
z^I \to e^{i (\theta - \hat{\theta})}
z^I \qquad \qquad b_\mu \to b_\mu
-  \partial_\mu \theta
+  \partial_\mu \hat{\theta}  \qquad \qquad c_\mu \to c_\mu
-  \partial_\mu \theta -  \partial_\mu \hat{\theta} \; .
\ee
Note that the last term in the action is gauge invariant because of
the Bianchi identity for $f$.

To this action, we can add a Lagrange multiplier term
\be
S_\sigma = \int d^3 x\, \frac{1}{8\pi} \sigma(x)
\epsilon^{\mu \nu \lambda} \partial_\mu f_{\nu \lambda}
\ee
and treat $f$ as an independent variable. The integral over $\sigma$
enforces the Bianchi identity. To be precise, we need $\sigma$ to be
periodic with period $2 \pi$. To see this, note that for monopole
configurations we can have, \footnote{In this theory, the minimum
monopole charge is double the one implied by the Dirac quantization
condition. We justify this in Appendix \ref{ChargeQuant}.}
\be
\label{quantcond}
\int d^3x\,\frac{1}{2} \epsilon^{\mu \nu \lambda}
\partial_\mu f_{\nu \lambda} =
\int_M dF = \int_{\partial M} F \in 4 \pi \mathbb{Z} \; .
\ee
For this, it is essential to note that $f$ is the sum of field
strengths for two gauge fields which are normalized conventionally (so
the gauge transformation is the derivative of an angle without any
numerical factors). So rather than a standard delta function, we want
a periodic delta function that allows these monopole
configurations. This is ensured by a $2 \pi$ periodicity of $\sigma$.

Starting from the combined action
\be
S = \int d^3 x
\left( - \big| \partial_\mu z^I +
i b_\mu z^I\big|^2 + \frac{k}{4 \pi}
\epsilon^{\mu \nu \lambda} b_\mu f_{\nu \lambda}
+ \frac{1}{8 \pi} \sigma \epsilon^{\mu \nu \lambda}
\partial_\mu f_{\nu \lambda}\right) \; ,
\ee
the equation of motion for $f$ gives
\be
b_\mu = \frac{1}{2k} \partial_\mu \sigma \; .
\ee
Using this, the full action reduces to
\be
\label{action2}
S = - \big| \partial_\mu z^I + \frac{i}{2k}
z^I \partial_\mu \sigma\big|^2 \; .
\ee
The gauge invariance transformation on $b$ translates to a gauge
invariance transformation on $\sigma$
\be
z^I \to e^{i \alpha(x)} z^I \qquad \qquad \sigma
\to \sigma - 2 k \alpha(x) \; .
\ee
We can now fix our gauge to set $\sigma = 0$. After doing this, we
still have residual gauge transformations
\be
\alpha(x) = \frac{\pi n}{k} \; ,
\ee
which leave $\sigma=0$. The moduli space is therefore characterized by
a set of eight complex numbers $z^I$, with gauge transformations that take
\be
\label{zkaction}
z^I \to e^{\pi i n/k} z^I
\ee
and
\be
z^I \to \bar{z}^I \; .
\ee
Here, the $\mathbb{Z}_2$ action and the $\mathbb{Z}_{2k}$ action don't
commute with each other for $k>1$, and the combined group is the
dihedral group $D_{2k}$. We conclude that the moduli space for level
$k$ is
\be
(\mathbb{R}^8 \times \mathbb{R}^8)/D_{2k} \; .
\ee
For $k=1$, this is just
\be
(\mathbb{R}^8 \times \mathbb{R}^8)/(\mathbb{Z}_2 \times \mathbb{Z}_{2}) \; ,
\ee
the moduli space of the superconformal theory that describes the
infrared physics of maximally SUSY $SO(4)$ Yang-Mills theory in 2+1
dimensions \cite{Sethi:1998zk}. In contrast, the superconformal theory
arising from $SU(3)$ gauge theory should have moduli space\footnote{In
general, the moduli space for gauge group $G$ with rank $n$ and Weyl
group ${\cal W}$ is $\mathbb{R}^{8n} /{\cal W}$.}
\be
(\mathbb{R}^8 \times \mathbb{R}^8)/S_3 \; .
\ee
For higher $k$, we conjecture that this theory describes the
low-energy physics of two M2-branes in M-theory with a generalized
orbifold action on $\mathbb{R}^8$. We expect that the geometrical
action is $\mathbb{R}^8/\mathbb{Z}_{2k}$. However, the orbifold group
must also act on the M2-brane fields, since the moduli space is not
just $(\mathbb{R}^8/\IZ_{2k})^2/\IZ_2$.

The orbifold in question should preserve 16 supersymmetries and
maximal $SO(8)$ R-symmetry for all $k$ and have the desired action
\eref{zkaction} on moduli space. However, except for $k=1,2$ there
appears to be no known orbifold with this property.\footnote{We thank
Nima Arkani-Hamed, Neil Lambert and David Tong for their comments on
this issue.}  The most supersymmetric singularities of the form
$\mathbb{R}^8/\IZ_{2k}$ in M-theory are
\cite{Halyo:1998pn,Morrison:1998cs}:
\beq
(z^1,z^2,z^3,z^4)\to (\omega z^1, \omega^{-1}z^2, \omega z^3,
\omega^{-1} z^4)
\eeq
where $\omega^{2k}=1$. Perhaps surprisingly, this action preserves as
many as 12 supersymmetries, or ${\cal N}=6$ in 3 dimensions, and also
gives rise to an R-symmetry $SU(4)\times U(1)$
\cite{Halyo:1998pn,Morrison:1998cs}. Even more intriguing, there are
two exceptions to this rule -- the cases with $k=1,2$. The former
obviously preserves ${\cal N}=8$, while the latter has also been
claimed to do the same \cite{Halyo:1998pn}. Though for general $k$
this orbifold does not appear to meet all the requirements, it is
possible that it actually preserves more supersymmetry and R-symmetry
than is apparent for reasons that we do not yet
understand.\footnote{This could be an M-theory analogue of the
mechanism in Ref. \cite{Duff:1998us} where a geometrical or
supergravity analysis yields misleadingly low amounts of supersymmetry
but additional stringy modes enhance the supersymmetry. In our system,
the fact that $\IZ_{2k}$ acts on the M2-branes may also be relevant.}

For the present, since we do not have a precise formulation
of the theory whose moduli space we have found, we simply think of it
as the theory of 2 M2-branes on an ``M-fold,'' and consider the $A_4$
theory at level $k$ to give a precise definition of the $\IZ_{2k}$
``M-fold''.

\section{Chiral primary operators}

In Ref. \cite{VanRaamsdonk:2008ft}, it was pointed out that it is
impossible to construct operators in the $A_4$ theory which
lie in tensor representations of $SO(8)$ with an odd number of
indices. This presented a puzzle for the interpretation of the $A_4$
theory as the worldvolume theory of a stack of M2-branes, since such
theories are believed (at least for three or more M2-branes) to have a
spectrum of chiral operators that includes these odd-indexed
representations. We will now see that with our proposed
interpretation, this is no longer a problem.

To see this most explicitly, let us focus on the case $k=1$ and
consider the UV gauge theory from which the superconformal field
theory flows. For the $SU(3)$ theory (or $SU(N)$ with $N>2$), the
scalar fields are seven Hermitian matrices, and we can construct
operators
\be
\label{ops}
{\rm STr}(X^{i_1} \cdots X^{i_n}) - {\rm SO(7) \; traces}
\ee
that should become a subset of the chiral primary operators in the
infrared limit (the others are generated by the $SO(8)$ rotations that
are not manifest in the UV).\footnote{Other trace structures give
additional operators. For $SU(2)$, such operators with an odd number
of $SO(7)$ indices do vanish identically, but for this case, the
moduli space is only eight dimensional.} On the other hand, in the
$SO(4)$ gauge theory, the scalars are antisymmetric matrices, so the
operators (\ref{ops}) vanish identically for odd numbers of
indices. This strongly suggests that the chiral primary operators with
odd numbers of $SO(8)$ indices will not be present in the infrared
theory either, so there is no conflict with identifying the IR limit
of the $SO(4)$ theory with the $k=1$ $A_4$ theory.

Generally, we expect that superconformal field theories that have the
same moduli space should also have the same spectrum of chiral
operators \cite{Bhattacharyya:2007sa}, so perhaps the discussion in
this section is somewhat redundant. However, it is interesting to
understand explicitly why the odd-indexed representations do not show
up in the $SO(4)$ case.

\section{Points on the moduli space}

In this section, we discuss more explicitly the connection between
points on the moduli space of the $A_4$ theory and
configurations of M2-branes on an orbifold. We begin with the simplest
case $k=1$. Here, the conjecture is that the moduli space
should coincide with the moduli space of two M2-branes on a
$\mathbb{Z}_2$ orbifold. This arises in the strong coupling limit of
type IIA string theory from a configuration of two D2-branes on an
$O2^-$ orientifold.

To understand the properties of such an orbifold, let us begin by
considering D2-branes at an $O2^-$ orientifold. The low-energy
worldvolume theory for these is $SO(4)$ maximally supersymmetric gauge
theory. The scalars in this theory are antisymmetric $4 \times 4$
matrices, and configurations for which the scalar potential vanishes
are gauge equivalent to
\be
X^i = \left( \ba{cccc} 0 & a^i & 0 & 0 \cr
-a^i & 0 & 0& 0 \cr 0 & 0 & 0 & b^i \cr 0 & 0 & -b^i& 0 \ea \right) \; .
\ee
Residual gauge transformations preserving this form allow us to make
the identifications
\be
(a^i, b^i) \equiv (b^i, a^i) \equiv (-a^i, -b^i) \equiv (-b^i, -a^i) \; ,
\ee
so the set of scalar field vevs with vanishing potential may be
described by the space $(\mathbb{R}^7 \times
\mathbb{R}^7)/(\mathbb{Z}_2 \times \mathbb{Z}_2)$.

The full moduli space of the IR limit of the $SO(4)$ gauge theory is
$(R^8 \times R^8)/(\IZ_2 \times \IZ_2)$, which we can describe by two
vectors in $\mathbb{R}^8$, subject to the identifications
\be
(A^I, B^I) \equiv (B^I, A^I) \equiv (-A^I, -B^I) \equiv (-B^I, -A^I) \; .
\ee
We can interpret $A$ and $B$ as the locations of the two
M2-branes. However, note that $(A^I, B^I)$ and $(A^I,-B^I)$ are
inequivalent, so the moduli space is not just a product of two
$\mathbb{R}^8/\mathbb{Z}_2$'s divided by the symmetric group, as one
might naively expect. In this characterization, a special role is
played by configurations $(A^I, A^I)$ where the branes are
coincident. These are invariant under the transformations
\be
(A^I, B^I) \equiv (B^I, A^I)
\ee
and so lie at special points of the moduli space.

Now, going back to the $A_4$ theory for $k=1$, we had derived
the moduli space as the space of complex vectors $z^I$ up to gauge
transformations
\be
z^I \to -z^I \qquad \qquad
z^I \to \bar{z}^I \; .
\ee
The first of these has no nontrivial fixed points, while the second
has a fixed point for real vectors. Thus, for our choice of gauge, it
is natural to make the associations
\bea
\label{zident}
Re(z^I) = A^I + B^I \nn\\
Im(z^I) = A^I - B^I
\eea
so that the fixed points of complex conjugation (equivalently, the
special points on the moduli space preserving $SU(2)$) are identified
with coincident branes. It may seem puzzling at first that there seem
to be more configurations than those with $z^I$ real that preserve
$SU(2)$ symmetry, namely any set of $z^I$ which lie in a line on the
complex plane. However, for these configurations, we do not have a
free abelian gauge field that can be dualized to a scalar, so these
are all gauge-equivalent to the configurations with $z^I$ real.

So far the discussion has dealt with $k=1$. It would be nice to carry
out a similar analysis for higher $k$, in particular to find the
precise relation between our coordinates $z^I$ on the moduli space and
the positions of the branes.

For $k>1$, we can also offer a rather heuristic geometrical
explanation for the origin of $D_{2k}$ as follows (we expect this
explanation could be made more precise with a better understanding of
the ``M-fold''). Suppose we bring two M2-branes to a $\IZ_{2k}$
orbifold. It is plausible that the result is, to start with, a theory
on $2k$ copies of the original branes, namely an $SU(2)^{2k}$ quiver
gauge theory. The quiver diagram is a $2k$-gon with the gauge fields
at the vertices. The form of the action \eref{markform} is consistent
with the presence of $2k$ $SU(2)$'s, except that at the origin of
moduli space the orbifold plane causes $k$ of these $SU(2)$'s to get
identified with each other, so that their action is $k$ times the
action of a single (level-1) $SU(2)$ Chern-Simons gauge
field. Likewise, the other $k$ $SU(2)$'s get identified and their
action is $k$ times that of another level-1 $SU(2)$ Chern-Simons gauge
field, appearing in the action with a negative sign. Given the quiver
interpretation, the associated symmetry group should be the set of all
discrete transformations that map the quiver to itself. This includes
cyclic rotations as well as reflections along any axis joining
opposite vertices. By definition, this is the group of symmetries of a
$2k$-gon, namely $D_{2k}$.

\section{Large $k$ and compactification}

In this section, we will see that the findings in
Ref. \cite{Mukhi:2008ux} fit very naturally with our interpretation. In
that paper it was found that expanding the $A_4$ action about a
special point on the moduli space where $SU(2)$ gauge symmetry is
preserved gives an action which is at leading order the maximally
supersymmetric $U(2)$ Yang-Mills theory. The extra $U(1)$ comes from
dualizing the scalar field that corresponds to multiplying all vevs by
a constant. This is not really a free scalar but is approximately free
at large distances from the orbifold, corresponding to the fact that
the theory on two M2-branes effectively has a centre-of-mass mode when
the branes are far away from the orbifold plane.

The procedure of \cite{Mukhi:2008ux} gives the Yang-Mills action plus
an infinite series of higher dimension operators. While the latter can
be decoupled in the limit $g_{YM}\to\infty$, the Yang-Mills action
simultaneously becomes strongly coupled in this limit. So there is no
limit where one really has finitely coupled D2-branes. However, the
the analysis of \cite{Mukhi:2008ux} was for level $k=1$. Repeating it
for general $k$, we find the following. By rescaling $X\to \sqrt{k}X,
\Psi\to \sqrt{k}\Psi$, we easily see that the action \eref{markform}
acquires an overall multiplicative factor of $k$. Denoting this scaled
action for the level-$k$ theory as $\cL^{(k)}$, we have
\beq
{\cal L}^{(k)} = k\,{\cal L}^{(k=1)} \; .
\eeq
Now in Ref. \cite{Mukhi:2008ux} the action ${\cal L}^{(k=1)}$ was
examined in the presence of a large vev $\langle X^{\phi(8)}\rangle =
v$ (there, this vev was called $R$ and later $g_{YM}$).  It was shown
(see Eq.(3.23) of that reference) that
\beq
{\cal L}^{(k=1)} = \frac{1}{v^2}\cL_0 + \frac{1}{v^3}\cL_1 + {\cal
O}\left(\frac{1}{v^4}\right)
\eeq
where $\cL_0$ is the action for an $N=8$ $SU(2)$ Yang-Mills theory.

For the Lagrangian $\cL^{(k)}$, we must define the Yang-Mills coupling
by
\be
\label{scalinglim}
g_{YM}^2 = \frac{v^2}{k} \; .
\ee
Taking the limit $k\to\infty,v\to\infty$ with $g_{YM}$ fixed, we see
that the Yang-Mills part of the action has a finite coupling
$g_{YM}$. However, successive terms scale to zero in this limit.
Therefore in the limit we obtain precisely the D2-brane
worldvolume theory with a tunable {\it finite} gauge coupling
$g_{YM}$, and no higher dimension operators.

With our interpretation of the theory, this observation is exactly
what we would expect. We have argued that points on the moduli space
preserving $SU(2)$ correspond to taking two coincident M2-branes away
from an orbifold fixed point. While a precise definition of the
orbifold action is lacking, for this discussion it is sufficient to
assume it leads to an opening angle that shrinks like $1/k$ as is the
case for standard orbifolds. Now, in the limit where $k \to \infty$,
the opening angle of the orbifold goes to zero, so at some point
sufficiently far out on the moduli space, the local geometry
approaches that of a cylinder $\mathbb{R}^7 \times S^1$. The scaling
limit below \eref{scalinglim} precisely takes the two M2-branes out
into this cylindrical space, where they should behave like two
D2-branes in type IIA string theory. So we expect a finitely coupled
$U(2)$ Yang-Mills theory -- and that is exactly what we find.

\section{Discussion}

In this paper, we have found that the moduli space for the
Bagger-Lambert $A_4$ theory at level $k$ is $(\mathbb{R}^8 \times
\mathbb{R}^8)/D_{2k}$, where $D_{2k}$ is the dihedral group of order
$4k$. Our interpretation is that the theory describes M2-branes on a
$\mathbb{Z}_{2k}$ ``M-fold,'' a generalization of the $\mathbb{Z}_2$
case defined by the uplift of the $O2^-$ orientifold in string theory.

We feel compelled to mention that the superconformal theories defined
as the IR fixed point of $U(2)$, $SO(4)$, $SU(3)$, $SO(5)$, and $G_2$
all have moduli spaces of the form
\beq
(\mathbb{R}^8 \times \mathbb{R}^8)/{\cal W}
\eeq
where ${\cal W}$ is respectively $D_1$, $D_2$, $D_3$, $D_4$, and
$D_6$. Within our interpretation, only the identification of the level
$k=1$ theory with $D_2$ seems natural, however, it may be that the
level $k=2$ and $k=3$ cases happen to coincide with the infrared limit
of $SO(5)$ and $G_2$ maximally supersymmetric gauge theory
respectively.\footnote{As we mentioned in Section 2, there is a
plausible maximally supersymmetric orbifold \cite{Halyo:1998pn} for
precisely $k=2$, suggesting a distinctive role for this case along
with $k=1$. The $k=2$ orbifold is related to the strong coupling limit of D2-branes at an $O2^+$ orientifold. A more detailed discussion of this case may be found in \cite{Lambert:2008et}.} This must be true unless there exist pairs of distinct $SO(8)$  superconformal field theories with the same moduli space.

The discussion in the limit of large-order $\IZ_{2k}$ orbifolds bears
a strong resemblance to the deconstruction approach to M5-branes
discussed in Ref. \cite{ArkaniHamed:2001ie}. In section IIIB of that
paper, a limit is taken where the order of the orbifold grows large
and simultaneously the D-branes are moved far away from the orbifold
so that effectively they end up propagating on a cylinder. It would be
interesting to explore whether the corresponding limit in our paper is
related to deconstruction and M5-branes.

\appendix
\section{Chern-Simons Level Quantization }\label{SpinCS}

The purpose of this appendix is to review the quantization of the
Chern-Simons level for the non-simply connected gauge groups,
$SO(n)$. None of the results are original but, particularly for the
spin Chern-Simons case, they are not as well-known as they should
be. We would like to thank Dan Freed for guiding us through the
computation below.

As shown by Dijkgraaf and Witten \cite{Dijkgraaf:1989pz}, the level of
a Chern-Simons theory with gauge group, $G$, is specified by an
element $a\in H^4(BG)$. If $H^4(BG)$ is one-dimensional, then $a$ is
an integer multiple of the generator, and the choice of Chern-Simons
action comes down to specifying that integer. When $G$ is not simply
connected, it is convenient to write the normalization of the
Chern-Simons action for $G$ \emph{relative} to that of the
Chern-Simons action for the simply connected covering group,
$\tilde{G}$. That is, the homomorphism $\tilde{G}\to G$ induces a map
$H^4(BG)\to H^4(B\tilde{G})$, and we might wish to note that the
generator of $H^4(BG)$ maps to some \emph{multiple} of the generator
of $H^4(B\tilde{G})$.

On a spin-manifold (which we certainly have, in our case, as we are
interested in supersymmetric theories), one can define a refined
version of Chern-Simons theory, called spin Chern-Simons
\cite{Jenquin-2005}. The precise definition of the action is slightly
more subtle (it involves the choice of a spin structure on the
3-manifold), but the \emph{variation} of the action, as one varies the
gauge connection is the same as for conventional Chern-Simons. The
only difference, from our perspective, is that the quantization
condition on the level is somewhat relaxed.

The level for spin Chern-Simons is specified by an element, $a\in
E^{4}(BG)$, of what Dan Freed calls \cite{Freed:2006mx}
$E$-cohomology, which combines information from the integer cohomology
with some mod-2 information. In particular, for any space, $X$, there
is a long exact sequence
\begin{equation}
\dots \to H^n(X) \to E^n(X) \to H^{n-2}(X,\mathbb{Z}/2)
\xrightarrow{\beta\circ Sq^2}H^{n+1}(X) \to E^{n+1}(X)\to\dots
\end{equation}
where the connecting homomorphism is the integer Bockstein, composed
with the second Steenrod square.

Let's apply this to the classifying spaces for $G=SO(n)$,
$\tilde{G}=Spin(n)$. For $n\geq 5$, $n=3$, $H^{4}(BSO(n))=\mathbb{Z}$,
with generator $p_{1}$. $H^{4}(BSO(4))=\mathbb{Z}\oplus \mathbb{Z}$,
with generators $p_{1}$ and
$e$. $H^{2}(BSO(n),\mathbb{Z}/2)=\mathbb{Z}/2$, with generator
$w_{2}$. The above long exact sequence gives rise to short exact
sequences
\begin{equation}
\begin{matrix}
0&\to& H^{4}(BSpin(n))&\xrightarrow{\simeq}& E^{4}(BSpin(n))&\to 0&&\\
&&\alpha\uparrow&&\gamma\uparrow&&&\\
0&\to& H^{4}(BSO(n))&\xrightarrow{\beta}& E^{4}(BSO(n))&\to&
H^{2}(BSO(n),\mathbb{Z}/2)&\to 0
\end{matrix}
\end{equation}

For $n\geq 5$, the map $\alpha$ is multiplication by 2 ($p_{1}$ for a
$Spin(n)$ bundle is always even), as is the map $\beta$. Hence the map
$\gamma$ is an isomorphism. Thus, for $SO(n)$, $n\geq5$, the
Chern-Simons coefficient $k$ must be even, while the spin Chern-Simons
coefficient can be any integer.

For $n=3$ the map $\alpha$ is multiplication by 4 (see, e.g., equation
(4.11) of \cite{Dijkgraaf:1989pz}). $\beta$ is still multiplication by
2, hence $\gamma$ is multiplication by 2. Thus, for ordinary $SO(3)$
Chern-Simons, $k\in 4\mathbb{Z}$, whereas for $SO(3)$ spin
Chern-Simons, $k\in 2\mathbb{Z}$.

For $SO(4)$, the case of actual interest in this paper, the map
$\alpha= \left(\begin{smallmatrix}1&1\\1&-1\end{smallmatrix}\right)$
has index-2. The map $\beta$ also has index-2. Hence $\gamma$ is an
isomorphism. Thus, for ordinary $SO(4)$ Chern Simons, the two $SU(2)$
levels satisfy $k_{L,R}\in\mathbb{Z}$, with $k_{L}-k_{R}\in
2\mathbb{Z}$. For $SO(4)$ spin Chern-Simons, $k_{L,R}$ can be any
integers.

As it turns, for the Bagger-Lambert $A_4$, we are interested in
$k_{L}= k$, $k_{R}=-k$, so the carry-away from this analysis is that
any integer value of $k$ is allowed, and there is no distinction
between the ordinary and spin Chern-Simons cases.

\section{Monopole Charge Quantization}\label{ChargeQuant}

In this appendix, we briefly explain why the quantization
(\ref{quantcond}) of the monopole charge in this theory is such that
the minimum charge is twice the one implied by Dirac's quantization
condition. It is well known that for 't Hooft-Polyakov monopoles, the
minimum charge is actually twice the Dirac value \cite{Montonen:1977sn}. In that
case, all fields transform in the adjoint representation of the gauge
group $SU(2)$, so effectively the gauge group is $SO(3)$.

In the theory we are considering, the gauge field $c_\mu$
corresponding to the unbroken $U(1)$ at generic points on the moduli
space sits inside the diagonal $SO(3) \in (SU(2) \times
SU(2))/\mathbb{Z}_2$, and all the matter fields transform in the
adjoint of this $SO(3)$, so the situation sounds similar to the 't
Hooft-Polyakov case. However, while monopole configurations in the 't
Hooft-Polyakov case are classified by
$\pi_2(SO(3)/SO(2))$,\footnote{In general, monopole solutions are
classified by $\pi_2(G/H)$ where $G$ is the gauge group and $H$ is the
unbroken subgroup.} monopole configurations in the BL theory should be
classified by elements of $\pi_2(SO(4)/SO(2))$. The embedding of the
diagonal $SO(3)$ in $SO(4)$ induces a natural map
\be
\label{embed}
\pi_2(SO(3)/SO(2)) \to \pi_2(SO(4)/SO(2)) \sim \mathbb{Z} \to \mathbb{Z}
\ee
but it is not obvious that this is an isomorphism. In particular, if
the generator of $\pi_2(SO(3)/SO(2))$ mapped to the square of the
generator of $\pi_2(SO(4)/SO(2))$, the BL theory would contain
monopoles with the minimal Dirac charge. It turns out that there can
be no such configurations, since the map (\ref{embed}) is onto, as we
will now show using the following theorem \cite{Bredon}.

{\bf Theorem}: If $p: Y \to B$ is a fibration and if
$y_0 \in Y$, $b_0 = p(y_0)$, and $F = p^{-1}(b_0)$, then
taking $y_0$ as the base point of $Y$ and of $F$ and $b_0$
as the base point of $B$, we have the exact sequence:
\[
\dots \rightarrow \pi_n(F) \rightarrow \pi_n(Y)
\rightarrow \pi_n(B) \rightarrow \pi_{n-1}(F) \rightarrow ...
\]

For our purposes, we take
\beas
Y &=& SO(4)/SO(2) = (SU(2) \times SU(2))/U(1) \cr
F &=& SO(3)/SO(2) = SU(2)/U(1) \cr
B &=& SU(2) \; .
\eeas
We can represent $Y$ by pairs $(U,V)$ of $SU(2)$ matrices where
\begin{equation*}
(U,V)\simeq (U e^{i\theta\sigma_{3}},Ve^{i\theta\sigma_{3}})\; .
\end{equation*}
 Take the fibration map $p$ to be
$(U,V) \to U V^{-1}$. We can take $y_0 = (1,1)$ so that $F$ is the
subgroup of $Y$ such that $U = V$, which is $SU(2)/U(1) =
SO(3)/SO(2)$. Now, a particular part of the exact sequence is
\[
\dots\to\pi_2(SO(3)/SO(2)) \xrightarrow{a} \pi_2(SO(4)/SO(2)) \rightarrow \pi_2(SU(2))\to\dots
\]
Since $\pi_2(SU(2)) = 0$, exactness implies that the map $a$ is onto, as we wanted to show.

\renewcommand\secstyle{\bfseries}

\section*{Acknowledgements}

We would like to thank Jim Bryan and Shiraz Minwalla for
helpful discussions. We thank Neil Lambert and David Tong for correspondence in sorting out the discrepancies between the original versions of our paper and of \cite{Lambert:2008et}.
J.D. would like to thank Dan Freed for topological guidance. The work of MVR has been supported in part by the Natural Sciences and Engineering Research Council of Canada, the
Alfred P. Sloan Foundation, and the Canada Research Chairs programme. The work of J.D.~was supported, in part, by NSF grant PHY-0455649.

\bibliographystyle{utphys}
\bibliography{m2-orb}

\providecommand{\href}[2]{#2}\begingroup\raggedright\begin{thebibliography}{10}

\bibitem{Bagger:2007jr}
J.~Bagger and N.~Lambert, ``Gauge symmetry and supersymmetry of multiple
  {M2}-branes,'' {\em Phys. Rev.} {\bf D77} (2008)  065008,
\href{http://arxiv.org/abs/0711.0955}{{\tt arXiv:0711.0955 [hep-th]}}.

\bibitem{Gustavsson:2007vu}
A.~Gustavsson, ``Algebraic structures on parallel {M2}-branes (v4),''
\href{http://arxiv.org/abs/0709.1260}{{\tt arXiv:0709.1260 [hep-th]}}.

\bibitem{Bagger:2007vi}
J.~Bagger and N.~Lambert, ``Comments on multiple {M2}-branes,'' {\em JHEP} {\bf
  02} (2008)  105,
\href{http://arxiv.org/abs/0712.3738}{{\tt arXiv:0712.3738 [hep-th]}}.

\bibitem{Mukhi:2008ux}
S.~Mukhi and C.~Papageorgakis, ``{M2 to D2},''
\href{http://arxiv.org/abs/0803.3218}{{\tt arXiv:0803.3218 [hep-th]}}.

\bibitem{Bandres:2008vf}
M.~A. Bandres, A.~E. Lipstein, and J.~H. Schwarz, ``{N = 8} superconformal
  {Chern--Simons} theories,''
\href{http://arxiv.org/abs/0803.3242}{{\tt arXiv:0803.3242 [hep-th]}}.

\bibitem{VanRaamsdonk:2008ft}
M.~Van~Raamsdonk, ``Comments on the {Bagger-Lambert} theory and multiple
  {M2}-branes,''
\href{http://arxiv.org/abs/0803.3803}{{\tt arXiv:0803.3803 [hep-th]}}.

\bibitem{Berman:2008be}
D.~S. Berman, L.~C. Tadrowski, and D.~C. Thompson, ``Aspects of multiple
  membranes,''
\href{http://arxiv.org/abs/0803.3611}{{\tt arXiv:0803.3611 [hep-th]}}.

\bibitem{Sethi:1998zk}
S.~Sethi, ``A relation between {N = 8} gauge theories in three dimensions,''
  {\em JHEP} {\bf 11} (1998)  003,
\href{http://arxiv.org/abs/hep-th/9809162}{{\tt arXiv:hep-th/9809162}}.

\bibitem{Dasgupta:1995zm}
K.~Dasgupta and S.~Mukhi, ``Orbifolds of {M}-theory,''
  \href{http://dx.doi.org/10.1016/0550-3213(96)00070-3}{{\em Nucl. Phys.} {\bf
  B465} (1996)  399--412},
\href{http://arxiv.org/abs/hep-th/9512196}{{\tt arXiv:hep-th/9512196}}.

\bibitem{Witten:1995em}
E.~Witten, ``Five-branes and {M}-theory on an orbifold,''
  \href{http://dx.doi.org/10.1016/0550-3213(96)00032-6}{{\em Nucl. Phys.} {\bf
  B463} (1996)  383--397},
\href{http://arxiv.org/abs/hep-th/9512219}{{\tt arXiv:hep-th/9512219}}.

\bibitem{Lambert:2008et}
N.~Lambert and D.~Tong, ``Membranes on an orbifold,''
\href{http://arxiv.org/abs/0804.1114}{{\tt arXiv:0804.1114 [hep-th]}}.

\bibitem{Morozov:2008cb}
A.~Morozov, ``On the problem of multiple {M2} branes,''
\href{http://arxiv.org/abs/0804.0913}{{\tt arXiv:0804.0913 [hep-th]}}.

\bibitem{Halyo:1998pn}
E.~Halyo, ``Supergravity on {AdS(5/4)} x {H}opf fibrations and conformal field
  theories,'' \href{http://dx.doi.org/10.1016/S0217-7323(00)00038-4}{{\em Mod.
  Phys. Lett.} {\bf A15} (2000)  397--406},
\href{http://arxiv.org/abs/hep-th/9803193}{{\tt arXiv:hep-th/9803193}}.

\bibitem{Morrison:1998cs}
D.~R. Morrison and M.~R. Plesser, ``Non-spherical horizons. {I},'' {\em Adv.
  Theor. Math. Phys.} {\bf 3} (1999)  1--81,
\href{http://arxiv.org/abs/hep-th/9810201}{{\tt arXiv:hep-th/9810201}}.

\bibitem{Duff:1998us}
M.~J. Duff, H.~Lu, and C.~N. Pope, ``{AdS(5) x S(5)} untwisted,''
  \href{http://dx.doi.org/10.1016/S0550-3213(98)00464-7}{{\em Nucl. Phys.} {\bf
  B532} (1998)  181--209},
\href{http://arxiv.org/abs/hep-th/9803061}{{\tt arXiv:hep-th/9803061}}.

\bibitem{Bhattacharyya:2007sa}
S.~Bhattacharyya and S.~Minwalla, ``Supersymmetric states in {M5/M2 CFTs},''
  \href{http://dx.doi.org/10.1088/1126-6708/2007/12/004}{{\em JHEP} {\bf 12}
  (2007)  004},
\href{http://arxiv.org/abs/hep-th/0702069}{{\tt arXiv:hep-th/0702069}}.

\bibitem{ArkaniHamed:2001ie}
N.~Arkani-Hamed, A.~G. Cohen, D.~B. Kaplan, A.~Karch, and L.~Motl,
  ``Deconstructing (2,0) and little string theories,'' {\em JHEP} {\bf 01}
  (2003)  083,
\href{http://arxiv.org/abs/hep-th/0110146}{{\tt arXiv:hep-th/0110146}}.

\bibitem{Dijkgraaf:1989pz}
R.~Dijkgraaf and E.~Witten, ``Topological gauge theories and group
  cohomology,''
\href{http://dx.doi.org/10.1007/BF02096988}{{\em Commun. Math. Phys.} {\bf 129}
  (1990)  393}.

\bibitem{Jenquin-2005}
J.~A. Jenquin, ``Classical {Chern-Simons} on manifolds with spin structure,''
  \href{http://arxiv.org/abs/math/0504524v2}{{\tt arXiv:math/0504524v2}}.

\bibitem{Freed:2006mx}
D.~S. Freed, ``Pions and generalized cohomology,''
\href{http://arxiv.org/abs/hep-th/0607134}{{\tt arXiv:hep-th/0607134}}.

\bibitem{Montonen:1977sn}
C.~Montonen and D.~I. Olive, ``Magnetic monopoles as gauge particles?,''
\href{http://dx.doi.org/10.1016/0370-2693(77)90076-4}{{\em Phys. Lett.} {\bf
  B72} (1977)  117}.

\bibitem{Bredon}
G.~E. Bredon, {\em Topology and Geometry}.
\newblock Springer Graduate Texts in Mathematics, 1993.

\end{thebibliography}\endgroup

\end{document}